# scientific reports

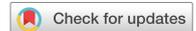

OPEN

# Entanglement propagation and dynamics in non-additive quantum systems

Guido Giachetti[1] & Nicolò Defenu[2]

The prominent collective character of long-range interacting quantum systems makes them promising candidates for quantum technological applications. Yet, lack of additivity overthrows the traditional picture for entanglement scaling and transport, due to the breakdown of the common mechanism based on excitations propagation and confinement. Here, we describe the dynamics of the entanglement entropy in many-body quantum systems with a diverging contribution to the internal energy from the long-range two body potential. While in the strict thermodynamic limit entanglement dynamics is shown to be suppressed, a rich mosaic of novel scaling regimes is observed at intermediate system sizes, due to the possibility to trigger multiple resonant modes in the global dynamics. Quantitative predictions on the shape and timescales of entanglement propagation are made, paving the way to the observation of these phases in current quantum simulators. This picture is connected and contrasted with the case of local many body systems subject to Floquet driving.

Many-body systems whose microscopic components interact via long-range power-law decaying potentials fall outside the traditional framework of complex systems, leading to rising wave of interest for their applications in quantum technology and computation[1–3]. Signatures characteristics of long-range physics are observed across several fields from astrophysics[4,5], to plasma physics[6] and experimental realizations on atomic, molecular and optical systems[3] (AMO), such as, trapped ions[1,7], Rydberg gases[8] and cold atoms into cavity experiments[2,9]. Plenty of exotic equilibrium and dynamical features have been detected, including dynamical phase transitions[10,11], time crystals[10,12,13], defect scaling[14,15] and long-lived metastable states[5]. These experimental results stimulated an impressive theoretical activity to characterize the equilibrium and dynamical critical scaling induced by long-range interactions in a wide variety of different systems[1–3]. Recently, the long-range research community has witnessed a rising interest in the dynamical behaviour of long-range systems due to their lack of equilibration[16–18]. This property is crucial to stabilize dynamical many-body phases such as time crystals[19–21].

Coherent many-body dynamics and collective out-of-equilibrium scaling, including thermalization and equilibration, can be described in terms of entanglement propagation, which can be also used to quantify the complexity of numerical simulations[22–27]. In the case of short-range interacting systems, a linear growth in time of bipartite von Neumann entropy, followed by a saturation to an equilibrium value, has been traced back to the presence of a finite speed information spreading and excitation propagation[23,28–30]. Slower growth has been observed, e.g. in case of ergodicity breaking[24,31–33].

Given the tight relation between entanglement scaling, excitations propagation and confinement[34–37], the breakdown of the locality concept induced by long-range interactions is expected to yield a strong impact on the traditional picture for correlations and entanglement spreading. The definition of long-range interactions generally refers to a two-body potential, which decays as a power-law of the distance $|\mathbf{i} - \mathbf{j}|$, i.e. $|\mathbf{i} - \mathbf{j}|^{-\alpha}$. As a function of $\alpha$ different situations arise: (i) for $\alpha > d$ long-range interactions can modify both the equilibrium critical properties[38] and the spreading of correlations[39,40] and entanglement[41], but they do not alter the fundamental properties of the system which, in general, thermalizates on a finite timescale[3,5]; (ii) on the contrary for $\alpha < d$, a novel regime appears, often called strong long-range[3]. There, fundamental concepts such as additivity of thermodynamic functions or the existence of a finite correlation length do not apply. As a consequence the description of entanglement spreading in terms of quasi-particle propagation is expected to completely breakdown. Accordingly, several numerical simulations have shown that the growth of entanglement entropy is dramatically slowed down[42–44], in spite of the fact that quantum correlations between far-apart degrees of freedom could build up quickly[45–47].

[1]SISSA and INFN Sezione di Trieste, Via Bonomea 265, 34136 Trieste, Italy. [2]Institut für Theoretische Physik, ETH Zürich, Wolfgang-Pauli-Str. 27, Zurich, Switzerland. ✉email: ndefenu@phys.ethz.ch

Here, the spreading of entanglement entropy after a sudden quench is investigated in a prototypical quantum long-range system. The entanglement features evidenced in previous studies is justified in terms of the peculiar spectral properties of strong long-range systems, which have been recently connected with the appearance of metastable states, whose lifetime grow with the system size[18]. The outcome of the present analysis provides analytical support to the suppression of entanglement spreading, evidenced in Refs.[42–44], while it yields solid indications of the existence of various novel phases at intermediate system sizes, i.e. the proper regime for quantum technological applications.

It is convenient to frame our studies in the context of quantum $O(n)$ rotor models, which constitute the prototypical tool in the context of quantum many-body physics. We consider a set of oscillator like variables $[\Phi_\mathbf{j}^a, \Pi_{\mathbf{j}'}^{a'}] = i\delta_{\mathbf{j},\mathbf{j}'}\delta_{a,a'}$ $(a = 1, \ldots, n)$, on a $d$-dimensional square lattice with $N = L^d$ sites, in presence of a generic quadratic interaction among different sites

$$H = \frac{1}{2}\sum_{\mathbf{j}}\left(|\vec{\Pi}_\mathbf{j}|^2 + r|\vec{\Phi}_\mathbf{j}|^2 + \frac{\lambda}{2n}|\vec{\Phi}_\mathbf{j}|^4\right) + \frac{1}{2}\sum_{\mathbf{j},\mathbf{j}'}t_{\mathbf{j}-\mathbf{j}'}|\vec{\Phi}_\mathbf{j} - \vec{\Phi}_{\mathbf{j}'}|^2. \tag{1}$$

In the following we will assume periodic boundary conditions.

An appropriate solution for the model dynamics after a global quench on any of the Hamiltonian parameters can be achieved, in the limit $n \to \infty$. There, the quartic term in Eq. (1) can be decoupled via the self-consistent relation

$$|\vec{\Phi}_\mathbf{j}|^4 \to \langle|\vec{\Phi}|^2\rangle|\vec{\Phi}_\mathbf{j}|^2. \tag{2}$$

Formally, this means that the correlation functions involving a finite number of components, at any time $t$, can be obtained via the decoupled theory up to $O(1/n)$ terms. In conclusion, the Hamiltonian in Eq. (1) at $n \to \infty$ can be replaced by its quadratic counterpart with a self-consistent effective mass

$$\mu(t) = r + \frac{\lambda}{2}\langle\Phi^2(t)\rangle. \tag{3}$$

The range of the interaction is encoded in the exact form of $t_{\mathbf{j}-\mathbf{j}'}$. The dynamical properties of the model in Eq. (1) have been deeply investigated for local and weak long-range couplings, presenting evidences of both dynamical phase transitions[48–50] and aging phenomena[51–53], which also occur at higher order in $1/n$[54]. Here, we are instead interested in the strong-long-range case, $t_{\mathbf{j}-\mathbf{j}'} = \frac{1}{N_\alpha}|\mathbf{j} - \mathbf{j}'|^{-\alpha}$ where $\alpha < d$ and $N_\alpha = \sum_\mathbf{j}|\mathbf{j}|^{-\alpha}$. The Kac scaling term $N_\alpha$ is needed[55] to ensure energy extensivity. In the following, it is convenient to focus on the $d = 1$ case, as the qualitative features of the evolution shall not depend on the dimension $d$.

The solution of the dynamics of the model passes through the introduction of the Fourier modes, labeled by the momentum $k = \frac{2\pi m}{N}$ with $m = -N/2 + 1, \ldots, N/2$. At variance with the traditional result in the nearest neighbours case (and with the weak long-range regime $\alpha > d$) the spectrum of the kinetic term remains discrete as the thermodynamic limit $N \to \infty$ is approached, see the Methods section "Spectrum". As a result, the dynamics of the system is reduced to the one of the effective model

$$H(t) = \sum_{k \geq 0}\left(\Pi_k^\dagger \Pi_k + (\omega_k^2 + \mu(t))\Phi_k^\dagger \Phi_k\right) \tag{4}$$

with frequency

$$\omega_k^2 = 1 - c_\alpha \int_0^{1/2} ds \frac{\cos(2\pi sm)}{s^\alpha}, \tag{5}$$

and $c_\alpha^{-1} = 2^{1-\alpha}(1-\alpha)$ so that $\omega_0 = 0$. For $|m| \gg 1$ instead the spectral levels accumulate around $\omega = 1$.

The Hamiltonian in Eq. (4), as well as its dynamics, is readily diagonalized in terms of the ladder operators $a_k, a_k^\dagger$, which represent the creation and annihilation operators of spin-waves of momentum $k$

$$\Phi_k = \frac{1}{\sqrt{2}}\left(f_k(t)a_k + f_{-k}^*(t)a_{-k}^\dagger\right), \tag{6}$$

$$\Pi_k = \frac{1}{\sqrt{2}}\left(\dot{f}_k^*(t)a_k^\dagger + \dot{f}_{-k}(t)a_{-k}\right). \tag{7}$$

The evolution equations of the amplitude functions $f_k(t)$ can be obtained via the Heisenberg equations of motion and read $\ddot{f}_k + (\mu + \omega_k^2)f_k = 0$. While the energy of any single mode is not conserved, we can define a conserved energy per particle, $\epsilon$, such that $\epsilon > \mu^2/2$ (see Methods section "Equations of motion" for further details).

The entire model dynamics is crucially dependent on the evolution of the effective mass $\mu(t)$. As shown in the Supplementary Material, Sec. 1, one can take $\mu(t)$ as an independent degree of freedom, and describe its evolution, together with the one of the $f_k$, in terms of the classical Hamiltonian

$$H = \frac{P_\mu^2}{2} + V(\mu) - \sum_k\left(\frac{|p_k|^2}{2m_k} + \frac{m_k}{2}(\mu + \omega_k^2)|f_k|^2\right), \tag{8}$$

where $P_\mu$ and $p_k$ are the conjugate momenta of $\mu$ and $f_k$, respectively. The potential, which regulates the evolution of the effective mass $\mu(t)$, reads $V(\mu) = (\mu - r)(\mu^2 - 2\epsilon) + 2(\mu - r)^2$. The effective masses $m_k$ are given by $m_k = \frac{4\lambda}{N}\mathcal{N}_k(1-\omega_k^2)$ with $\mathcal{N}_k = 1 + \langle n_k \rangle + \langle n_{-k} \rangle$, $n_k$ being the initial occupation number of each mode. See Ref.[56] for an analogous picture in the classical case. Let us notice that the minus sign in Eq. (8) does affect the equation of motions for $f_k$. The corresponding equation of motion for $\mu$ can instead be written as

$$\ddot{\mu} = -V'(\mu) + \frac{1}{2}\sum_k m_k |f_k|^2 \qquad (9)$$

the last term, that we denotes with $g(t)$, encodes the contribution of the other modes. Because of the properties of the spectrum, $Nm_k \to 0$ for $k \gg 1$ and $g(t)$ is negligeble in the thermodynamic limit as long as $n_k, f_k \sim O(1)$. More precisely, as shown in the Methods section "Spectrum", $g(t) \sim N^{-\zeta}$, with $\zeta = \max(1, 2-2\alpha)$ so that, in the short time regime, $\mu(t)$ evolves as the position of a fictitious particle in the potential $V(\mu)$, with the corresponding energy conservation $\mathcal{E} = \frac{1}{2}\dot{\mu}^2 + V(\mu)$. Despite the cubic nature of the potential $V(\mu)$ one can prove that $\mathcal{E} < 0$ for any viable initial condition in the system, see Supplementary Material Sec. 1. As a consequence, the motion of the effective mass in the accessible region is a periodic oscillation around the only minimum of the potential $V(\mu)$.

The aforementioned classical dynamics only applies as long as the external force $g(t)$ remains negligible, this condition may be violated even at short timescales in two cases:

(1) The initial state at $t = 0$ contains at least one macroscopically populated mode leading to some extensive $n_k$.
(2) The external drive $g(t)$ may become extensive due to one (or more) $k$ modes undergoing a parametric resonance.

Scenario (1) occurs for any dynamics beginning in the magnetized ground-states at $r < -\lambda/2$, where the $k = 0$ mode acquires a macroscopic occupation, see Supplementary Material Sec. 2. Even for an initially negligible external drive $g(t)$, −, the oscillations of the effective mass $\mu(t)$ may act as a drive for one or more modes, leading to scenario (2). Indeed, the equation of motion for each of the $f_k$ is in the form of a Hill equation, and the Floquet theorem implies that one can always find a pair of oscillatory independent solutions, $f_k^\pm(t+T) = e^{\pm i\mu T} f_k(t)$ with $\mu > 0$, or a resonant pair, $f_k^\pm(t+T) = e^{\pm \kappa T} f_k(t)$ with $\kappa > 0$ (see Methods section "Floquet theorem for the Hill equation" for a derivation of this result). In the latter case, the corresponding excitation undergoes a resonance and the mode occupation $\sim e^{2\kappa t}$ will scale as $\sim O(N)$, signaling the rise of many-body correlations in the system.

Thus, the time-scale for the external drive $g(t)$ to become relevant is reduced to $t_q \sim \log N$, which has to be compared with $t_q \sim N^\zeta$ in the classical phase. In summary, while suppressed in the strict thermodynamic limit, for any finite size $N$ the spreading of quantum correlations may still occur on a time-scale of order $N$ or $\log N$ in the classical and resonant phase respectively, see Fig. 1.

Given the importance that long-range interactions are ought to play in quantum information and technology applications, it is natural to focus on regime (2), where quantum correlations can arise even for initial states in the disordered phase. The actual boundaries of the mesoscopic quantum phases in the phase space of the initial parameters cannot be simply determined analytically, so that it is convenient to resort to numerical analysis. As the Hamiltonian (8) is an integral of motion, the $k$-th mode cannot resonate as long as $\mu(t) + \omega_k^2 > 0$ for every $t$ (see also Supplementary Material Sec. 1). In particular, for $r > 0$, no resonance can occur, in agreement with the fact that the initial model in Eq. (1) lies in the symmetric phase. For $r < 0$, resonant phases can actually be observed and a rich phase diagram arises, as shown in Fig. 2 for the case $r = -1$, $\alpha = 0.5$ (see Methods section "Numerical methods" for details on the numerics). By decreasing further $\epsilon$, large oscillations begin triggering the resonance in the $k = 0$ mode, then the $m = 1$ mode start resonating as well as the corresponding $\mu(t) + \omega_k^2$ is no longer positive definite, followed by the $m > 1$ ones. Let us notice that the presence of a gapped spectrum is crucial in order to have regions in parameter space with only a finite number of resonances.

As a consequence, we get an effective description of the system in terms of a finite number of degrees of freedom by neglecting the non-resonant excitations in Hamiltonian (8). This is the case shown in Fig. 3, where the single particle picture (orange line) is shown to fail on a time-scale $t_q \sim \log N$, while a two mode approximation (cyan line) faithfully reproduces the numerical results (dotted blue line). On a timescale $t > t_q$ the resonant mode grows large, and then becomes again negligible, leading to periodic oscillations, punctured with periodic bursts, corresponding to the resonances.

As the number of resonant modes increases, the behavior of $\mu(t)$ becomes increasingly involved. Let us notice, however, that for thermal-like initial conditions, $\mathcal{N}_k, f_k(0), p_k(0)$ will depend only on the $\omega_k$, so that the $m \gg 1$ modes will behave as a single particle.

### Entanglement growth

Our overall scenario can be confirmed by the study of entanglement scaling after a quench in the bare mass parameter $r$ ($r^- \to r^+$) at $t = 0$, assuming the system is in the ground state for $t < 0$ in the disordered phase, where the above analysis holds. As a measure of the entanglement it is convenient to choose the Von Neumann entropy relative to the partition of the chain in two intervals. Due to the factorisation of the rotor models interaction term at large-$n$, and to the fact that the initial state is Gaussian, one can apply the formalism developed in Ref.[57] for computing the entanglement entropy (see Methods section (Entanglement Entropy for the finite interval)) starting from the momentum and position two-point functions.

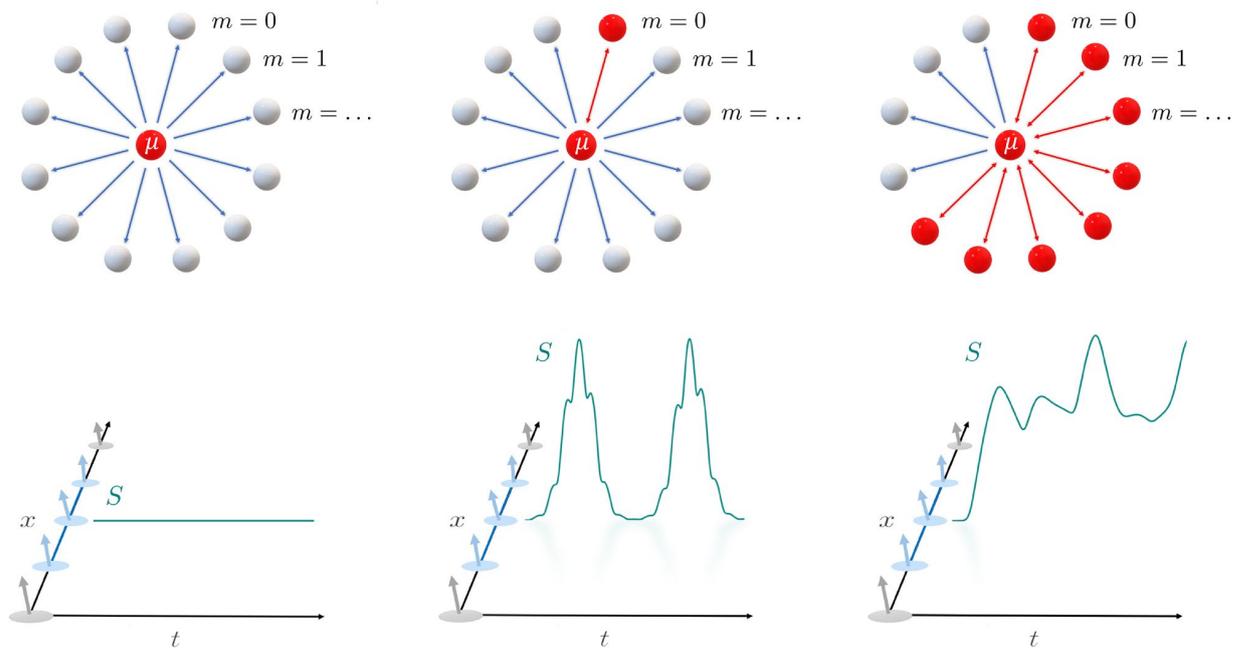

**Figure 1.** Schematic depiction of the phases of our model along with the behavior of the bipartite Von Neumann entropy $S(t)$. *Left*: the classical phase, in which the classical dynamics of $\mu(t)$ is not influenced by the presence of the quantum bath up until $t_q \sim N^\zeta$, so that there is no production of entanglement. *Center*: the resonant phase, in which the $m=0$ mode becomes resonant and affect the dynamics on the timescale $t_q \sim \log N$, causing periodic bursts in $S(t)$. *Right*: the multi-resonant phase, in which a larger number of modes resonate, causing a more complex oscillatory behavior with a finite production of entanglement.

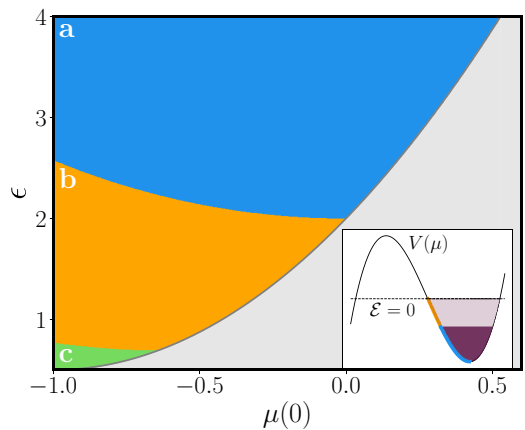

**Figure 2.** Phase diagram of the model for $r=-1, \alpha=0.5$ as a function of the energy per particle $\epsilon$ and the initial effective mass $\mu(0)$, assuming $\dot\mu(0)=0$. Only the region with $\mu(0)$ between $r$ and the minimum of $V(\mu)$ is shown, being all the other initial conditions nonphysical or redundant. In blue we have the resonance-free region (in which $\mu(t)$ is periodic); in orange the region in which the mode $k=0$ is resonant; in green the region, where multiple modes above $k=0$ are resonant as well. *Inset*: the potential $V(\mu)$ for $\epsilon=2.25$, in which the values of $\mu(0)$ corresponding to different phases and the relative values of the classical energy $\mathcal{E}$, see Eq. (8), are outlined.

Let us, then, consider the growth of entanglement entropy in an interval of length $\ell$. For simplicity, we first restrict to the case in which only the $k=0$ mode may be resonant. Given that the single particle spectrum for long-range interactions accumulates at high-energy in the $N \to \infty$ limit[18] at order $1/N$ one can discard the $k$ dependence of all modes at $k>0$. Therefore, we can replace $f_k(t)$ and $\dot f_k(t)$ with their high energy limit $f_\pi(t)$,

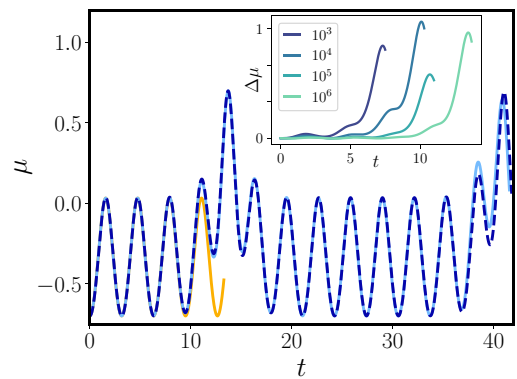

**Figure 3.** Behavior of $\mu(t)$ with $r = -1$, $\lambda = 1.24$, $N = 10^6$ for the initial conditions $\mu(0) = -0.7$, $\dot\mu(0) = 0$, $\epsilon = 1.2$ (dotted blue line) compared with the corresponding single-particle classical picture (continuous orange) and the two-particle one (continuous cyan line). Since for this initial conditions only the $k = 0$ mode is resonant, the latter reproduces the right evolution, while the former fails on a finite timescale. *Inset:* difference between $\mu(t)$ and the single-particle picture for different values of $N$, showing how the time-scale on which the approximation breaks down grows as $\log N$.

---

$\dot f_\pi(t)$. As a result, the dynamical theory reduces to the one of a classical particle coupled to the resonant $k = 0$ mode plus a $\sim N$ times degenerate high energy mode, leading to the simplified quantum correlations

$$\begin{aligned}
\langle \Phi_x(t)\Phi_{x'}(t)\rangle &= \frac{\delta_{x'x}}{2}|f_\pi(t)|^2 + \frac{1}{2N}|f_0(t)|^2, \\
\langle \Pi_x(t)\Pi_{x'}(t)\rangle &= \frac{\delta_{x'x}}{2}|\dot f_\pi(t)|^2 + \frac{1}{2N}|\dot f_0(t)|^2, \\
\langle \Phi_x(t)\Pi_{x'}(t)\rangle &= \frac{\delta_{x'x}}{2}\bigl(\mathrm{Re}(f_\pi(t)\dot f_\pi^*(t)) + i\bigr) \\
&\quad + \frac{1}{2N}\mathrm{Re}(f_0(t)\dot f_0^*(t)).
\end{aligned} \quad (10)$$

It is worth noting that the latter expressions approximate the actual correlations of the system up to $O(N^{-1})$ terms, which may become relevant if the length of the considered interval $\ell$ becomes of order $N$.

The resulting expression for the Von Neumann entropy of an interval of length $\ell \ll N$ (see Supplementary Material Sec. 3)

$$S(t) = \left(\sqrt{1+\ell\Delta(t)}+1\right)\ln\frac{\sqrt{1+\ell\Delta(t)}+1}{2} \\
- \left(\sqrt{1+\ell\Delta(t)}-1\right)\ln\frac{\sqrt{1+\ell\Delta(t)}-1}{2}, \quad (11)$$

where

$$N\Delta(t) = |f_\pi(t)\dot f_0(t)|^2 + |f_0(t)\dot f_\pi(t)|^2 \\
- 2\mathrm{Re}\bigl(f_0(t)\dot f_0^*(t)\bigr)\mathrm{Re}\bigl(f_\pi(t)\dot f_\pi^*(t)\bigr). \quad (12)$$

As expected, the $1/N$ factor included in the definition of $\Delta(t)$ suppresses the propagation of entanglement unless the $k = 0$ mode is resonant. Thus, the parametric resonance results in the production of entanglement on an intermediate time-scale $t \sim \log N$, which shall be accessible even for large many-particle experiments. The behaviour of $S(t)$ in the resonant phase is shown in Fig. 4. $S(t)$ is characterized by periodic bursts on a time-scale $t_q \sim \log N$, after which it returns to its initial value. Apart from lack of equilibration, which is signalled by the $S(t)$ function not saturating, the periodic generation and fading of entanglement is a peculiar feature of long-range interactions, which is evidenced here for the first time. Moreover, Eq. (11) implies that, the larger $\ell$ is, the faster quantum correlations will be established. This is in antithesis with the case of local and weak long-range systems, where the short-time growth of entanglement is independent on $\ell$, due to the light-cone like structure of the Lieb-Robinson bound[58].

The impact of the number of resonant modes (and thus of effective degrees of freedom) on the evolution of $S(t)$ is apparent if we consider the case multiple resonant modes beyond the $k = 0$ one. In this case, the analytic expression for the correlation functions in Eq. (11) does not apply. However, following Ref.[57] one can derive the evolution of $S(t)$ in the multi-resonant phase, see Fig. 4.

In summary, while in the $N \to \infty$ limit, non-additive interaction hinders the spreading of quantum excitations, for large but finite $N$, the out-of-equilibrium phase diagram of long-range interacting quantum systems presents a rich mosaic of phases. In the prototypical case of $O(n)$ field theories, we argued that such phases may be characterized by the scaling of the entanglement entropy. In particular, for sudden quenches from any ground state in the disordered equilibrium-phase, we identify three phases:

(a) *Classical phase:* in the thermodynamic limit, the many-body dynamics corresponds to the one of a single classical particle, representing the effective mass $\mu(t)$. Correspondingly, at finite sizes, no entanglement emerges in the system up to a scale $t_q \sim N^\zeta$.

(b) *Resonant phase:* for larger initial energies the oscillations of the effective mass $\mu(t)$ trigger a resonance in the lowest $k = 0$ mode. As a consequence, the entanglement entropy grows in quasi-periodic bursts, the first of which occurs at a time-scale $t_q \sim \log N$. Yet, following each burst, the amount of entanglement vanishes and classical dynamics is restored.

(c) *Multi-resonant phase:* at high energy, the effective mass oscillations generated by the instantaneous quench are strong enough to generate multiple resonances, leading to a mosaic of different phases, each characterised by a number of active modes in the dynamics. While the phenomenology of all these phases is analogous to the single resonance one, the entanglement does no longer come back to zero. Indeed, as a large number of resonance is triggered, see Fig. 4, a finite amount of entanglement is generated at long-times, which is however smaller than its short-range counterpart. In spite of such apparent saturation of entanglement entropy, the quasi-periodic oscillations persist at any large time, so that no actual equilibration occurs even in this case.

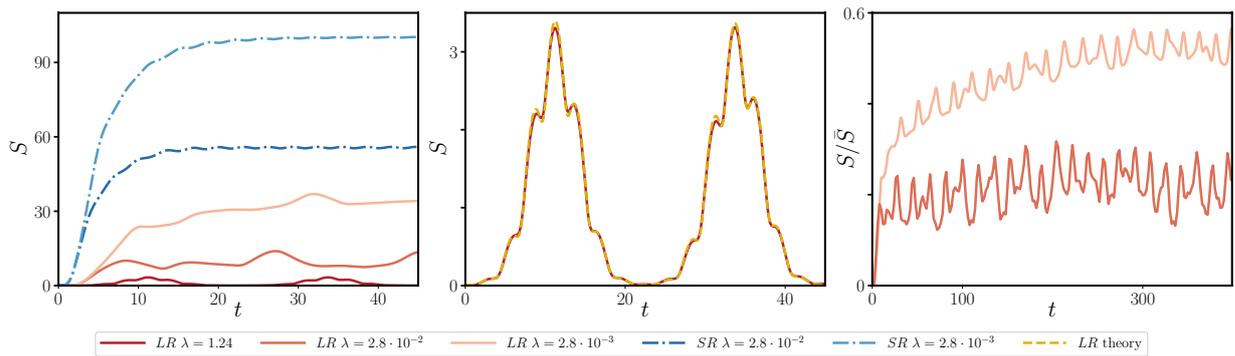

**Figure 4.** *Left panel:* behavior of the Von Neumann entropy $S(t)$ relative to a $\ell = 10$ interval after a ground state quench of $r$ from 1 to $-1$, for $N = 10^4$. The dark red, red and orange curves correspond to different values of $\lambda$ ($\lambda = 1.24$, $\lambda = 0.028$, $\lambda = 0.0028$), which in turn corresponds to have 1, 116 and 221 resonant modes respectively. All curves refer to the case $\alpha = 0.5$. The dotted dark blue and blue curves correspond to the short-range counterpart ($\alpha = \infty$) of the $\lambda = 0.028$, $\lambda = 0.0028$ cases. *Central panel:* Behavior of $S(t)$ in the single-resonance phase (red), characterized by periodic bursts, compared with the analytical prediction of Eq. (11) (dashed yellow line). *Right panel:* Production of entanglement for larger times in the multi-resonant phase, compared with the correspondent limiting value $\bar{S}$ for the same quench in the short-range limit ($\alpha \to \infty$). See Methods section "Numerical methods" for a more detailed explanation of the numerical procedure.

The present picture can be traced back to the discreteness of the single particle spectrum evidenced in Ref.[18]. Then, the appearance of resonant phases belongs to the rich landscape of peculiar phenomena caused by strong long-range interactions, including metastable states[16,17], ensemble inequivalence[59] or lack of adiabaticity[60,61]. In this perspective, our picture shall be regarded as universal and will be observed in any systems where long-range couplings generate a discrete spectrum, including cavity QED and trapped ions experiments. As a further evidence of the universal nature of our findings, the slow growth of entanglement entropy, which we explicitly connected to the presence of a discrete spectrum, has also been found in numerical simulations of the ferromagnetic long-range Ising model after a sudden quench[44]. While this slow growth of entanglement may be confused with the one found in finite range and weak long-range systems[41], the latter is generated by quasi-particle confinement[34] or Bloch oscillations[62] and, due to its prethermal character, can only be observed at short times. On the contrary, the current picture follows from the collective character of strong long-range interactions, which produce an effective global coupling between the degrees of freedom in the thermodynamic limit, see Fig. 1, stabilising the aforementioned phases even at large times.

Along the same lines, the collective character of long-range interactions with $\alpha < d$ prevents the divergence of low-energy excitations and stabilises the $1/n$ approach employed to study the dynamics, making it trustworthy also in the thermodynamic limit. As a result, the effective mass $\mu(t)$ acts in all respects as a global drive and generates the resonant phases in analogy with previous studies of bosonic theories subject to an external periodic force[63]. Nevertheless, in the present case, the back action of the resonant modes on the drive prevents the linear growth of the entropy with time. This is consistent with the isolated nature of the system and, consequently, with the absence of any external energy source.

Our investigations belong to a long strand of literature that studies universal dynamical phenomena in flat interacting systems, starting from the celebrated Lipkin-Meshkov-Glick and Sachdev-Ye-Kitaev models[64–66] and extending up to astrophysical scales[67]. Our findings represent a substantial step forward in this direction as they encompass in a single framework the cases of both flat ($\alpha = 0$) and of fully-connected power-law decaying interactions ($0 < \alpha < d$). Thus, we have presented a general framework capable to predict and justify a large variety of experimental and numerical observations, arising in experimental realisation of strong long-range interactions[1–3]. The present description exploits two main traits of strong long-range systems, namely the discreteness of the spectrum[18] and the global collective coupling emerging from their collective character, see the role of the global effective mass $\mu$ in Eq. (8).

Here, a first application of this general framework to sudden quenches and entanglement propagation has been presented, resulting in an explicit mapping to periodically driven systems (see the Hill equation in the Methods section "Floquet theorem for the Hill equation") which has no counterpart in short- or weak long-range interacting systems. This previously unknown relation allowed us to fully characterise the dynamical phase diagram of strong long-range quantum systems and produce several analytical estimations over the scaling of the proper time of each phase, which may be used to justify and guide experimental endeavours.

In conclusion, the entanglement propagation in strong long-range interacting quantum systems is suppressed in the thermodynamic limit, but a rich phenomenology appears at intermediate system sizes. The strong collective character of long-range interactions, see Fig. 1, explicitly relates the quench dynamics with the one of a periodically driven quantum system, yielding a robust description for entanglement propagation and dynamics. In particular, the resonant phases (b) and (c) display a logarithmic scaling of time with the size $t_q \sim \log N$, allowing their observation also in mesoscopically and, possibly, macroscopically large quantum simulators. In contrast with the known picture for local and, in general, additive systems, these phases do not have pre-thermal character, but they persist at all times, preventing decoherence and, possibly, allowing more robust technological applications.

## Methods

**Spectrum.** Let us now derive Eq. (5). If we write down the kinetic part of the Hamiltonian in terms of the Fourier modes we have:

$$\frac{1}{2} \sum_{j,j'} t_{j-j'} (\Phi_j - \Phi_{j'})^2 = \sum_k \omega_k^2 \Phi_k^\dagger \Phi_k \tag{13}$$

where

$$\omega_k^2 = \sum_{r \neq 0} t_r (1 - \cos kr). \tag{14}$$

Now, being $t_r = N_\alpha^{-1} |r|^{-\alpha}$, with $N_\alpha = \sum_{r \neq 0} |r|^{-\alpha}$, we can write $\omega_k^2 = 1 - \tilde{t}_k / \tilde{t}_0$ where

$$\tilde{t}_k = \sum_{r=1}^{N/2} r^{-\alpha} \cos kr. \tag{15}$$

Since $k = \frac{2\pi}{N} m$, we notice that, as $N \to \infty$

$$\tilde{t}_k = N^{1-\alpha} \sum_{r=1}^{N/2} N^{-1} \left(\frac{r}{N}\right)^{-\alpha} \cos \frac{2\pi m r}{N}$$
$$= N^{1-\alpha} \int_0^{1/2} ds \frac{\cos(2\pi m s)}{s^{-\alpha}} + O(1). \tag{16}$$

Being $\tilde{t}_0 = N^{1-\alpha} \int_0^{1/2} ds\, s^{-\alpha} + O(1) = N^{1-\alpha} c_\alpha + O(1)$, we get the expression (5) up to $O(N^{\alpha-1})$ corrections.

Let us now consider the spectral average of a regular function $G(x)$, namely

$$\langle G(\omega^2) \rangle = \frac{1}{N} \sum_k G(\omega_k^2). \tag{17}$$

Since, from Eq. (14),

$$\sum_k \tilde{t}_k = 0, \qquad \sum_k \tilde{t}_k^2 = N \sum_{r \neq 0} r^{-2\alpha}, \tag{18}$$

at the lowest non trivial order

$$\langle G(\omega^2) \rangle \approx G(1) + \frac{G''(1)}{\tilde{t}_0^2} \sum_{r=1}^{N/2} r^{-2\alpha}. \tag{19}$$

Depending whether $1/2 < \alpha < 1$ or $0 < \alpha < 1/2$ the correction is then of order $N^{-1}$ or $N^{-2+2\alpha}$ respectively. Let us notice how this is completely different to what happens for the nearest-neighbours case, in which the gap between any two adjacent energy levels vanishes as $N \to \infty$. On the contrary, the result in Eq. (16) is reminiscent of the $\alpha = 0$ case, where, due to full permutational symmetry, $\omega_k = 1 - \delta_{k,0}$. The $\alpha < d$ case retains the discreteness of the well known $\alpha = 0$ case, but the finite excitations energy allow for the appearance of multiple resonances in the out-of-equilibrium dynamics.

**Equations of motion.** From Eq. (4) one can derive the Heisenberg equations of motion for the operators $\Pi_k, \Phi_k$

$$\dot{\Phi}_k = \Pi_k^\dagger, \qquad \dot{\Pi}_k = -(\omega_k^2 + \mu(t))\Phi_k^\dagger \qquad (20)$$

In particular, being the evolution equation linear, it is convenient to expand $\Phi_k$ and $\Pi_k$ as

$$\Phi_k = \frac{1}{\sqrt{2}}\left(f_k(t)a_k + f_{-k}^*(t)a_{-k}^\dagger\right),$$
$$\Pi_k = \frac{1}{\sqrt{2}}\left(\dot{f}_k^*(t)a_k^\dagger + \dot{f}_{-k}(t)a_{-k}\right) \qquad (21)$$

where $a_k, a_k^\dagger$ and $a_{-k}, a_{-k}^\dagger$ are two independent sets of ladder operators which do not depend on time. The canonical commutation relations introduce the constraints $f_k(t) = f_{-k}(t)$, $\text{Im}(f_k^* \dot{f}_k) = 1$. Let us notice that, in terms of the $f_k$, we can express the equal-time correlators of the model as

$$\langle \Phi_k(t)\Phi_{k'}(t)^\dagger \rangle = \frac{\mathcal{N}_k}{2}|f_k(t)|^2 \delta_{kk'},$$
$$\langle \Pi_k(t)\Pi_{k'}(t)^\dagger \rangle = \frac{\mathcal{N}_k}{2}|\dot{f}_k(t)|^2 \delta_{kk'}, \qquad (22)$$
$$\langle \Phi_k(t)\Pi_{k'}(t) \rangle = \frac{\delta_{kk'}}{2}\left(\mathcal{N}_k \text{Re}(f_k(t)\dot{f}_k^*(t)) + i\mathcal{S}_k\right),$$

where $\mathcal{N}_k = 1 + \langle n_k \rangle + \langle n_{-k} \rangle$, $\mathcal{S}_k = 1 + \langle n_k \rangle - \langle n_{-k} \rangle$, $\langle n_k \rangle = \langle a_k^\dagger a_k \rangle$ being the *initial* occupation number of the corresponding mode. If we use the definitions of Eq. (6) into Eqs. (20) we get for the $f_k$

$$\ddot{f}_k(t) + \left(\omega_k^2 + \mu(t)\right)f_k(t) = 0. \qquad (23)$$

Once written in terms of the real and imaginary part of $f_k$, one can interpret this as the equations of motion for a set of two-dimensional isotropic harmonic oscillators. The frequency varies in time, and it is determined self consistently by Eq. (3). The evolution respects the constraints $\text{Im}(f_k^* \dot{f}_k) = 1$, as they correspond to the conserved angular momenta of the oscillators. The single oscillator energy is instead not conserved. However, one can verify that the energy per particle

$$\epsilon = \frac{\lambda}{2N}\sum_k \mathcal{N}_k(|\dot{f}_k|^2 + \omega_k^2|f_k|^2) + \frac{1}{2}\mu^2 \qquad (24)$$

is a constant of motion. Let us notice that, by definition $\mu > r$ and $\mu^2 < 2\epsilon$ must hold.

**Floquet theorem for the Hill equation.** We remind briefly the basics of the Floquet theory. Let us notice that, in case $\mu(t)$ is periodic, the equation of motion (23) for the single mode has the form of an Hill equation

$$\ddot{f}(t) + a(t)f(t) = 0, \qquad (25)$$

with $a(t) = a(t+T)$. We may now consider the two independent solutions $f_1(t)$ and $f_2(t)$ such that $f_1(0) = 1$, $\dot{f}_1(0) = 0$ and $f_2(0) = 0$, $\dot{f}_2(0) = 1$ respectively. Let us notice that the Wronskian of the solutions $W = f_1\dot{f}_2 - f_2\dot{f}_1$ of the system is such that $\dot{W} = 0$. Now we notice that, being $a(t)$ periodic of period $T$, $f_1(t+T), f_2(t+T)$ can be seen as a new pair of independent solutions, and thus expressed as a linear combination of $f_1(t), f_2(t)$ namely

$$\begin{pmatrix} f_1(t+T) \\ f_2(t+T) \end{pmatrix} = C \begin{pmatrix} f_1(t) \\ f_2(t) \end{pmatrix}, \qquad (26)$$

where $C$ is a constant square matrix of order 2. In particular, imposing $W(t) = W(t+T)$ we find $\det C = 1$. Let us now consider two independent linear combinations $f_\pm(t)$, of $f_{1,2}(t)$ such that

$f_\pm(t+T) = \Lambda_\pm f_\pm(t)$, $\Lambda_\pm$ being the eigenvalues of $C$. On the other hand, since $C$ is real and $\det C = 1$ we have that only two cases are possible, either $\Lambda_\pm = e^{\pm\kappa T}$ or $\Lambda_\pm = e^{\pm i\mu T}$ for some real (positive) $\kappa, \mu$. While in the former case we have a quasi-periodic solution, in the latter we have a parametric resonance.

**Entanglement entropy for the finite interval.** Following the procedure of [57,68], we introduce the matrix of the correlations

$$\gamma = \text{Re}\begin{pmatrix} \langle \Phi_x(t)\Phi_{x'}(t) \rangle & \langle \Phi_x(t)\Pi_{x'}(t) \rangle \\ \langle \Pi_x(t)\Phi_{x'}(t) \rangle & \langle \Pi_x(t)\Pi_{x'}(t) \rangle \end{pmatrix}. \qquad (27)$$

The Von Neumann entropy $S(t)$ can be expressed in terms of the symplectic spectrum $\{\sigma_n\}$ of $\gamma_{\text{red}}$, i.e. the matrix reduced to the subspace $x, x' \in (0, \ell)$. The symplectic spectrum is defined such that $\{\sigma_n^2\}$ is the spectrum of $-(J\gamma_{\text{red}})^2$, $J$ being the symplectic unity

$$J = \begin{pmatrix} 0 & \mathbb{I}_\ell \\ -\mathbb{I}_\ell & 0 \end{pmatrix}. \tag{28}$$

In terms of the $\sigma_n$, we have

$$S = \sum_n s(\sigma_n), \tag{29}$$

where:

$$s(\sigma) = \left(\sigma + \frac{1}{2}\right) \ln\left(\sigma + \frac{1}{2}\right) - \left(\sigma - \frac{1}{2}\right) \ln\left(\sigma - \frac{1}{2}\right). \tag{30}$$

**Numerical methods.** In order to preserve the symplectic structure of the dynamics, the equation of motions for $\mu(t)$ and for the $f_k$ have been integrated with a second-order, volume-preserving, algorithm for separable Hamiltonian. The time-step chosen wad $\Delta t = 0.05$. In Fig. 2 we have numerically determined the matrix $C$ in Eq. (26) for each $k$, and verified the stability conditions Tr $C < 2$, which guarantees imaginary eigenvalues. In Fig. 4, the procedure of section "Floquet theorem for the Hill equation" was carried out numerically, by computing the matrix $\gamma$ and its symplectic spectrum for different values of $t$.

### Data availability
Data sharing not applicable to this article as no datasets were generated or analysed during the current study.

### Acknowledgements
This research was funded by the Swiss National Science Foundation (SNSF) grant number 200021 207537 and by the Deutsche Forschungsgemeinschaft (DFG, German Research Foundation) under Germany's Excellence Strategy EXC2181/1-390900948 (the Heidelberg STRUCTURES Excellence Cluster). G.G. thanks Giuseppe di Giulio and Sara Murciano for the useful discussions, and Alessandro Santini for the insightful comments.

### Competing interests
The authors declare no competing interests.

### Additional information

**Supplementary Information** The online version contains supplementary material available at https://doi.org/10.1038/s41598-023-37984-3.

**Correspondence** and requests for materials should be addressed to N.D.

**Reprints and permissions information** is available at www.nature.com/reprints.

**Publisher's note** Springer Nature remains neutral with regard to jurisdictional claims in published maps and institutional affiliations.

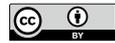 **Open Access** This article is licensed under a Creative Commons Attribution 4.0 International License, which permits use, sharing, adaptation, distribution and reproduction in any medium or format, as long as you give appropriate credit to the original author(s) and the source, provide a link to the Creative Commons licence, and indicate if changes were made. The images or other third party material in this article are included in the article's Creative Commons licence, unless indicated otherwise in a credit line to the material. If material is not included in the article's Creative Commons licence and your intended use is not permitted by statutory regulation or exceeds the permitted use, you will need to obtain permission directly from the copyright holder. To view a copy of this licence, visit http://creativecommons.org/licenses/by/4.0/.

© The Author(s) 2023

# Supplementary Material

## 1 Evolution of $\mu(t)$

Let us now derive the equation of motion for the effective mass $\mu(t)$. By tacking the time derivative of both sides of Eq. (3) of the main text

$$\dot{\mu} = \frac{\lambda}{2N}\sum_k \mathcal{N}_k \dot{f}_k f_k^* + c.c.$$

$$\ddot{\mu} = \frac{\lambda}{2N}\sum_n \mathcal{N}_k \left(|\dot{f}_n|^2 + \ddot{f}_n f_n^* + c.c.\right) \tag{1}$$

$$= \frac{\lambda}{N}\sum_k \mathcal{N}_k |\dot{f}_k|^2 - \frac{\lambda}{N}\sum_k \mathcal{N}_k \left(\mu + \omega_k^2\right)|f_k|^2 ,$$

from which, exploiting the conservation of energy and Eq. (3) of the main text, we can write

$$\ddot{\mu} = 2\epsilon + 2r\mu - 3\mu^2 - 2(\mu - r) + g(t) \tag{2}$$

where we introduced

$$g(t) = \frac{1}{2}\sum_k m_k |f_k(t)|^2, \tag{3}$$

and $m_k = 4\lambda \mathcal{N}_k(1-\omega_k^2)/N$. Let us notice that this equation, along with the equation of motion for the $f_k$, can be derived from the Hamiltonian (9) of the main text

$$H = \frac{P_\mu^2}{2} + V(\mu) - \sum_k \left(\frac{|p_k|^2}{2m_k} + \frac{m_k}{2}(\mu + \omega_k^2)|f_k|^2\right). \tag{4}$$

Now, $g(t) \sim \langle 1-\omega_k^2\rangle \sim O(N^{-\zeta})$ as long as $f_k = O(1)$, so that its contribution to the equations of motion of $\mu(t)$ is negligible. In this limit, we can thus set $f_k = 0$ in the above Hamiltonian and consider the single particle dynamics:

$$H = \frac{P_\mu^2}{2} + V(\mu). \tag{5}$$

Within the same approximation

$$\epsilon = \frac{\lambda}{2N}\sum_k \mathcal{N}_k |\dot{f}_k|^2 + \mu - r + \frac{1}{2}\mu^2 \tag{6}$$

up to $N^{-\zeta}$ correction. Let us notice that, the motion of this effective particle takes place within the bounded region of the potential, since the corresponding (conserved) energy $\mathcal{E} = \frac{\dot{\mu}^2}{2} + V(\mu)$ can only be negative. Indeed, by exploiting the Cauchy-Schwartz inequality, we find

$$\dot{\mu}^2 = \left(\text{Re}\left(\frac{\lambda}{N}\sum_k \mathcal{N}_k \dot{f}_k f_k^*\right)\right)^2 \leq \left|\frac{\lambda}{N}\sum_k \mathcal{N}_k \dot{f}_k f_k^*\right|^2$$

$$\leq \left(\frac{1}{N}\sum_k \mathcal{N}_k |f_k|^2\right)\left(\frac{1}{N}\sum_k \mathcal{N}_k |\dot{f}_k|^2\right). \tag{7}$$

Now, putting together Eq. (6) and Eq. (3) of the main text and we find the constraint

$$\dot{\mu}^2 \leq 2\left(2\epsilon - 2(\mu - r) - \mu^2\right)(\mu - r) = -V(\mu), \tag{8}$$

from which the condition $\mathcal{E} < 0$ follows.

Let us now consider the modes $f_k = O(1)$. In this case the conserved energy becomes

$$\mathcal{E} = \frac{P_\mu^2}{2} + V(\mu) - \sum_k \frac{m_k}{2}\left(|\dot{f}_k|^2 + (\mu + \omega_k^2)|f_k|^2\right), \tag{9}$$

which differs from the single-particle energy for a quantity of order $N^{-\zeta}$. Now, if along the single-particle trajectory $\mu + \omega_k^2 > 0$ for every $k$, then the curve defined in the space of parameters $f_k, \dot{f}_k, \mu, \dot{\mu}$ remains close to the $f_k = 0$ trajectory: in this case then, no resonance is possible. In particular, since $\mu(t) > r$, this ensures that no resonance actually occurs in the $r > 0$ case. The same reasoning leads to the the condition $r + \omega_k^2 > 0$ to prevent the resonance of modes with $k > 0$.

## 2 Derivation of the ground state properties

Let us now derive the expression for $\mu_{\text{gs}}$ and $\epsilon_{\text{gs}}$. Since each oscillator $\Phi_k$ is now in the ground state, we have

$$\langle \Phi_k(0)\Phi_{k'}(0)^\dagger \rangle = \frac{1}{2\sqrt{\omega_k^2 + \mu_{\text{gs}}}} \, \delta_{k'k},$$
$$\langle \Pi_k(0)\Pi_{k'}(0)^\dagger \rangle = \frac{1}{2}\sqrt{\omega_k^2 + \mu_{\text{gs}}} \, \delta_{k'k}, \tag{10}$$
$$\langle \Phi_k(0)\Pi_{k'}(0) \rangle = \frac{i}{2} \, \delta_{k'k}$$

from which, exploiting Eqs. (22) of the Methods, we find

$$f_k(0) = \left(\omega_k^2 + \mu_{\text{gs}}\right)^{-1/4}, \quad \dot{f}_k(0) = i\left(\omega_k^2 + \mu_{\text{gs}}\right)^{1/4}, \tag{11}$$

valid up to an immaterial phase factor. Since $\mu_{\text{gs}}$ is now a positive constant, the solution of Eqs. (23) of the Methods is given by

$$f_k(t) = \left(\omega_k^2 + \mu_{\text{gs}}\right)^{-1/4} e^{i\sqrt{\omega_k^2 + \mu_{\text{gs}}}\, t}. \tag{12}$$

Finally, from Eq. (3),

$$\mu_{\text{gs}} = r + \frac{\lambda}{2N} \sum_k \frac{1}{\sqrt{\omega_k^2 + \mu_{\text{gs}}}}. \tag{13}$$

In our case we can replace $\sqrt{\omega_k^2 + \mu_{\text{gs}}}$ with $\sqrt{1 + \mu_{\text{gs}}}$, up to $O(N^{-\zeta})$ corrections, obtaining:

$$\mu_{\text{gs}} = r + \frac{1}{2}\frac{\lambda}{\sqrt{1 + \mu_{\text{gs}}}}. \tag{14}$$

This always has a unique solution. Since we are implicitly assuming $\mu_{\text{gs}} > 0$, in order to have an oscillatory behavior for the $k=0$ mode, we have to require $r > -\lambda/2$. For $r < -\lambda/2$ the $k=0$ mode acquires a non-zero occupation number, signalling the emergence of a finite magnetization. The fact that the system undergoes a phase transition even in one dimension is not surprising, since the Mermin-Wanger theorem no longer holds in presence of long-range interactions. The corresponding energy per particle is, up to $O(N^{-\zeta})$ correction, given by

$$\epsilon_{\text{gs}} = \frac{\lambda}{2N}\sum_k \frac{1}{\sqrt{\omega_k^2 + \mu_{\text{gs}}}}\left(2\omega_k^2 + \mu_{\text{gs}}\right) + \frac{1}{2}\mu_{\text{gs}}^2 = \frac{1}{2}\frac{\lambda}{\sqrt{1 + \mu_{\text{gs}}}}\left(2 + \mu_{\text{gs}}\right) + \frac{1}{2}\mu_{\text{gs}}^2. \tag{15}$$

This allows for a simple physical interpretation: indeed $\epsilon_{\text{gs}}$ is such that $V'(\mu_{\text{gs}}) = 0$, so that the ground state corresponds to the stable equilibrium for the fictitious particle in the potential $V(\mu)$.

## 3 Von Neumann entropy for a single resonance

We now apply the procedure exposed in Methods. In our case, from Eq. (10) of the Methods we have

$$\gamma_{\text{red}} = \frac{1}{2}\begin{pmatrix} Q(t) & R(t) \\ R(t) & P(t) \end{pmatrix}, \tag{16}$$

where $Q(t), P(t), R(t)$ are $\ell$ by $\ell$ matrices defined as

$$Q(t) = |f_\pi(t)|^2 \mathbb{I}_\ell + \frac{\ell}{N}|f_\pi(t)|^2 \mathbb{P},$$
$$P(t) = |\dot{f}_0(t)|^2 \mathbb{I}_\ell + \frac{\ell}{N}|\dot{f}_\pi(t)|^2 \mathbb{P}, \tag{17}$$
$$R(t) = \text{Re}\left(f_\pi(t)\dot{f}_\pi^*(t)\mathbb{I}_\ell + \frac{\ell}{N}f_0(t)\dot{f}_0^*(t)\mathbb{P}\right),$$

with $\mathbb{P}_{j,k} = \frac{1}{\ell}, \ \forall\, j,k = 1,\ldots \ell$. Since $[P, R] = 0$ and $[Q, R] = 0$ we find

$$-(J\gamma_{\text{red}})^2 = \frac{1}{4}\begin{pmatrix} PQ - R^2 & 0 \\ 0 & PQ - R^2 \end{pmatrix}. \tag{18}$$

Using the fact that $\mathrm{Im}(f_k(t)\dot{f}_k^*(t)) = 1$ and $\mathbb{P}^2 = \mathbb{P}$ we have

$$PQ - R^2 = \mathbb{I} + \ell\Delta(t)\mathbb{P} + \ell^2 N^{-2}\,\mathbb{P}, \tag{19}$$

with $\Delta(t)$ of Eq. (12) of the main text. For a finite interval, the last term on the r.h.s. of Eq. (19) is negligible while the second one may become $O(1)$ on a timescale $t_q \sim \ln(N)$. Then all the eigenvalues of $-(J\gamma_{\mathrm{red}})^2$ are $\frac{1}{4}$ but two which are

$$\frac{1}{4} + \frac{\ell}{4}\Delta(t). \tag{20}$$

The symplectic spectrum is finally given by

$$\begin{aligned}\sigma_1 = \sigma_2 &= \frac{1}{2}\sqrt{1+\ell\Delta(t)}, \\ \sigma_n &= \frac{1}{2} \quad \forall\, n = 3, \ldots 2\ell.\end{aligned} \tag{21}$$

Substituting in Eq. (29) of the Methods, we notice that only the first two eigenvalues, coming from the resonant mode, do actually contribute to $S(t)$, and we recover the expression (11) in the Main text.